# Vibrational Frequency used as Internal Clock Reference to access Molecule - Metal Charge Transfer Times

P. Jakob* and S. Thussing

Fachbereich Physik und Wissenschaftliches Zentrum für Materialwissenschaften der Philipps-Universität Marburg, Renthof 5, 35032 Marburg, Germany

* email:  peter.jakob@physik.uni-marburg.de

**Abstract**

Dynamical charge transfer processes at molecule-metal interfaces proceed in the few fs time scale that renders them highly relevant to electronic excitations in optoelectronic devices. Yet, knowledge thereof is limited when electronic ground state situations are considered that implicate charge transfer directly at the fermi energy. Here we show that such processes can be accessed by means of vibrational excitations, with non-adiabatic electron-vibron coupling leading to distinct asymmetric line shapes. Thereby the characteristic time scale of this interfacial dynamical charge transfer can be derived by using the vibrational oscillation period as an internal clock reference.

**Introduction**

In modern material sciences, molecule - metal interfaces play a decisive role in the functionality of molecular electronic devices. Despite their importance in technological applications our microscopic understanding is lagging behind, probably due to the experimental difficulties to detect and single out the weak interface signatures. Acquiring comprehensive and conclusive evidence to expand our knowledge base regarding charge transfer processes at the molecule - metal interface remains challenging, though, especially defining the dynamics at ultrafast time scales [1].

Electron transfer processes across an interface connecting two weakly coupled electronic systems can be accessed in various ways, e.g. by means of breaking junctions [2], scanning tunneling microscopy [3], or by using photoemission spectroscopy, often in combination with pump-probe techniques [4-6]. While the former provide a precise map of the charge transport through well-defined single molecule junctions, the dynamics and time frame of the very process is revealed by the latter only. In photoemission experiments excitation typically involves electronic states far above $\varepsilon_F$; adoption of the derived dynamical properties to the neutral ground state (or the-like) might then be problematic. Even more so, it is the charge transport at (or close to) zero energy, i.e. right at the Fermi energy, that is most relevant to the functioning of molecular electronic circuitry. To assess the role of the molecule - metal interface in charge transport within devices it is thus indispensible to focus on charge transfer processes at or close to $\varepsilon_F$.

In this Letter we will explore the electron dynamics for ground-state configurations of molecular adsorbates by means of vibrational mode excitation, provoking charge oscillations between the molecule and the metal substrate. Strictly speaking, this approach comprises a non-ground state situation as well (vibrational excitation); however, deviation thereof is only minor, representing a key advantage with respect to the mentioned electron spectroscopies. In our analysis we will take advantage of the fact that the degree of non-adiabaticity in the electron - vibron coupling is directly reflected in the vibrational line shape. As a key ingredient we will be using the vibrational frequency of specific modes of the molecular constituents to serve as an intrinsic and precise clock to extract the relevant time scale. The usage of an internal clocking scheme is both compelling and effective and has been used in a different context in core level photoemission studies (core hole clock) to derive charge transfer times down to the sub fs range for adsorbed rare gas atoms and diatomic molecules [7, 8].

We will be examining vibrational line shapes to address the topic of electron - phonon/vibron coupling in general, as well as questions regarding adiabaticity in particular. Electron - vibron coupling describes the oscillatory motion of atoms within (adsorbed) molecules in the context of their coupling to electronic levels of the molecule and the metal substrate; given that the Born-Oppenheimer approximation is satisfied, the coupling between these two entities will proceed adiabatically. As vibrational excitations represent a non-equilibrium situation, e.g. in view of the electronic response of the surroundings, non-adiabaticity, to some extent, is commonly expected.

Historically, non-adiabatic electron - vibron coupling, in combination with electron - hole pair excitation has been introduced to explain short lifetimes of vibrational modes associated with (small) molecules (CO, NO) adsorbed on metal substrates [9]. This irreversible process must be distinguished from the reversible charge oscillations across the molecule-metal interface (interfacial dynamical charge transfer, IDCT). Essentially, the former process primarily accounts for the homogeneous line broadening, studied in various instances in the past, while the latter produces asymmetric line shapes of the respective vibrational bands; these can be accessed experimentally and have been described theoretically by Persson and Persson [9, 10], as well as Langreth [11], who have derived expressions for the associated Fano-type line shapes. The model system at that time, adsorbed CO, however, did not show a noteworthy asymmetry, pointing at a negligible deviation from adiabaticity in the electron - vibron coupling. For quite some time, H(D) on W(100) was the only system displaying pronounced asymmetric vibrational line shapes [12, 13].

Molecular layers studied in the present work are more promising as they come with a well-defined interface separating the two electronic systems, molecule and metal substrate; moreover, they display a characteristic set of vibrational modes with specific symmetry properties and displacement patterns. Infrared absorption spectroscopy (IRAS), used in this work, represents a versatile method to study such systems; due to its high spectral resolution and strict symmetry selection rules identification and characterization of molecular species is straightforward, based on their vibrational signatures (see Supplemental Material for a detailed description. This is essential insofar as molecule - metal charge transfer processes can be uniquely correlated with a particular molecular species.

**Experimental**

Experiments were performed in an ultrahigh vacuum chamber (base pressure p = 5×10$^{-11}$ mbar) which contained facilities for fourier-transform infrared absorption spectroscopy (FT-IRAS), spot-pro le analysis low energy electron diffraction (SPA-LEED) and thermal desorption spectroscopy (TDS). All infrared spectra (spectral range of 600 - 4000 cm$^{-1}$) were taken at a surface temperature of 80 K, using an evacuated Bruker IFS 66v/s instrument and a liquid N$_2$ cooled MCT (HgCdTe) detector; spectral resolution typically was 2 cm$^{-1}$ and about 1000 scans have been co-added. Characterization of molecular layers regarding long range ordering is achieved using an Omicron SPA-LEED system. For thermal desorption experiments as well as examination of evaporant cleanliness, a quadrupole mass spectrometer (Pfeiffer, QMG 700) with mass range 0 - 1024 u was used. Temperatures have been measured using a K-type thermocouple, (laser)welded to the edges of the Ag(111) or Au(111) single crystals. The samples were mounted to a liquid He or N$_2$ cooled cryostat and could be heated resistively with a linear heating rate.

The various molecular species have been deposited by means of home-made evaporators. The sample temperature T$_{sample}$ during deposition typically was 300 K; alternatively, growth has been carried out at T$_{sample}$ ≈ 80 K, with adequate post-deposition annealing to 300 - 500K to ensure thermal equilibration and long range ordering. The evaporator temperatures were adjusted within ±0.1 K to yield deposition rates of about 0.2 monolayers per minute. Specifically, temperatures have been adjusted to: 310 K (Tetracene), 390 K (NTCDA), 530 K (PTCDA), and 500 K (CuPc). During deposition the background pressure remained below 1×10$^{-10}$ mbar. For Tetracene, extra precautions (permanent cooling to 273 K) had to be implemented to lower its vapor pressure in order to avoid unnecessary exposure of the UHV chamber. Before conducting the experiments, the sample was generally cleaned by Ar$^+$ ion sputtering (700 eV, 1 µA, T$_{sample}$ = 380 K, Δt = 30 min), followed by annealing to 780 K for 5 - 10 min.

**Results and Discussion**

In Figure 1, a compilation of IR spectra associated with various molecular species adsorbed on noble metal substrates is presented. Specifically, Tetracene, PTCDA, NTCDA and CuPc, all of them planar molecules, have been deposited on Ag(111) or Au(111) surfaces. According to the literature, all molecular species are adsorbed with their molecular plane oriented parallel to the surface [14-17]; deviations from planarity due to interactions with the substrate are only moderate or weak [15, 17]. This weak molecule - metal coupling strength is reflected in insignificant frequency shifts of vibrational modes with respect to the values of the respective isolated or bulk species. Noteworthy exceptions are the C=O stretch modes of the acyl groups at the corners of NTCDA and PTCDA which are subject to a bold red shift by about 150 cm$^{-1}$ when adsorbed on Ag(111), due to chemical interactions with substrate Ag atoms [17, 18]. The key observation when comparing individual spectra in Figure 1 is that the intensities of in-plane modes (ν > 1000 cm$^{-1}$) vary distinctly, while the IR absorptions of out-of-plane modes (prevailing at ν < 1000 cm$^{-1}$) display about similar dipolar strength. This marked variation of in-plane mode intensities is ascribed to IDCT being active for the bottom three spectra, while it is absent for the two uppermost spectra.

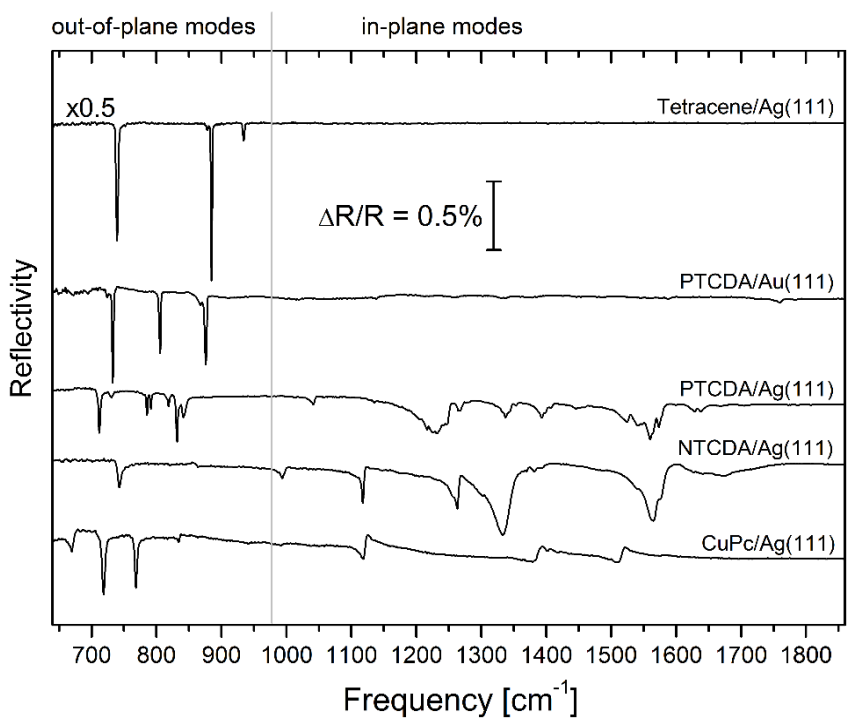

**Figure 1:**
Collection of infrared absorption spectra, obtained from CuPc/Ag(111), NTCDA/Ag(111), PTCDA/Ag(111), PTCDA/Au(111), and Tetracene/Ag(111) monolayers. The spectra were obtained at 80 K using a spectral resolution of 2 cm$^{-1}$; individual curves are vertically offset for better display.

Apparently, the strength or weakness of in-plane vibrational modes is neither a specific property of the substrate, nor of a particular molecular species, as IDCT is absent for PTCDA on Au(111) while it is quite prominent when adsorbed on Ag(111). On Ag(111), IDCT is likewise observed for adsorbed NTCDA and CuPc, besides PTCDA, but not for Tetracene. A prerequisite for IDCT to be operative is that molecular electronic states at or close to the Fermi energy $\varepsilon_F$ exist [9, 19, 20]. In accordance with this hypothesis, studies of the electronic structure have identified features at or slightly below $\varepsilon_F$ ascribed to the former lowest unoccupied molecular orbital (f-LUMO) for PTCDA, NTCDA, and CuPc on Ag(111) [21 - 24]; for Tetracene/Ag(111) and PTCDA/Au(111) the weaker adsorbate-substrate interaction renders the respective LUMO empty [25, 26].

Partially filled molecular orbitals may form as a result of hybridization with electronic levels of the substrate, leading to a static charge transfer between molecule and substrate; more importantly, such a constellation of electronic states enables charge oscillations across the molecule - metal interface induced by vibrational mode excitations.

To refine our understanding of IDCT we will take a closer look at the various contributions to a vibrational mode's dynamic dipole moment $\mu_{dyn}$. For adsorbed species the contributions to $\mu_{dyn}$ essentially comprise two components: (i) $\mu_{mol}$ coming from nuclear motion of atoms within a molecule carrying (static) charges, in addition to contributions connected with charge redistribution within the molecule, and (ii) $\mu_{IDCT}$ associated with dynamic charge transfer across the adsorbate - substrate interface. Electron - vibron coupling and dynamical charge flow within the molecule is expected to proceed adiabatically. $\mu_{mol}$ must then be discriminated from $\mu_{IDCT}$, as coupling to the substrate, in general, is much weaker; this means that it is more difficult to maintain the charge equilibrium between metal and molecule at short time scales, leading to a delayed response of $\mu_{IDCT}$ with respect to $\mu_{mol}$.

The effect of such a delay in the built-up of the dynamic dipole moment and its interaction with incoming IR radiation has been described by Langreth by defining $\mu_{IDCT}$ and accordingly $\mu_{dyn} = \mu_{mol} + \mu_{IDCT}$ as a complex quantity [11]. $\mu_{dyn}$ can then be split in a purely real $\mu_{mol}$ and a complex term $\mu_{IDCT} = \mu_{1,IDCT} + i\mu_{2,IDCT}$, so that $\mu_{dyn} = (\mu_{mol} + \mu_{1,IDCT}) + i\mu_{2,IDCT}$. Here, the real part refers to the instantaneous response of the dynamical dipole moment. The complex nature of $\mu_{IDCT}$ or $\mu_{dyn}$ then accounts for the delayed response of the interfacial charge flow with respect to vibrational motion. The molecular layers presented in Figure 1 are ideal model systems as their weak adsorbate - substrate interaction leads to a significant phase lag between $\mu_{mol}$ and $\mu_{IDCT}$, resulting in vibrational line shapes with pronounced asymmetry.

To quantify the degree of non-adiabaticity, Langreth [11] has introduced a parameter $y = \omega\tau$, describing the ratio of the imaginary and the real part of the complex quantity $\mu_{dyn}$, that is, $\omega\tau = \frac{\mu_{2,IDCT}}{\mu_{mol}+\mu_{1,IDCT}}$. A related quantity $\omega\tau_{IDCT}$ refers to $\mu_{IDCT}$ alone instead of $\mu_{dyn}$ so that $\omega\tau_{IDCT} = \frac{\mu_{2,IDCT}}{\mu_{1,IDCT}}$. For a particular vibrational mode at $\omega_0$ with narrow linewidth $\gamma \ll \omega_0$, absorption is non-negligible in the region of the resonance only and we can take $y = \omega\tau \cong \omega_0\tau$. We then obtain an expression that combines the charge transfer time $\tau_{IDCT}$ with the asymmetry parameter $\omega_0\tau$ of the resulting Fano-type line shapes through $\tau_{IDCT} = \frac{1}{\omega_0}\left(\frac{\mu_{mol}}{\mu_{1,IDCT}} + 1\right)\omega_0\tau$. In fact, the sign of $\mu_{IDCT}$ need not be identical to $\mu_{mol}$ when considering different types of vibrational modes. This leads to a partial compensation of $\mu_{IDCT}$ and $\mu_{mol}$ and gives rise to negative values of $\omega_0\tau$ even though $\tau_{IDCT}$, as a rule, will always be positive.

In Figure 2 enlarged sections of the IR spectra of CuPc/Ag(111) and of NTCDA/Ag(111) are shown. The displayed frequency regions contains various in-plane as well as out-of-plane vibrational modes. We find that the respective asymmetry parameter values differ notably for the two types of modes; in fact, they consistently show a reversal of their asymmetry. Specifically, CuPc in-plane modes ($a_{1g}$) at 671 and 835 cm$^{-1}$ display positive asymmetry parameters, while out-of-plane modes ($a_{2u}$) at 719 and 768 cm$^{-1}$ yield negative $\omega_0\tau$. Similarly, the asymmetry changes its sign for the NTCDA modes located at 742 and 862 cm$^{-1}$, both representing out-of-plane modes ($b_{3u}$), as compared to those at 995 and 1118 cm$^{-1}$ which have been identified as in-plane modes ($a_g$). We conclude that rather than the particular vibrational frequency, it is the specific character of a vibrational mode that decides on the sign of the respective asymmetry parameter.

By applying the Langreth expression [11], the asymmetry parameter $\omega_0\tau$ is readily retrieved by means of curve fitting. We have analyzed various in-plane vibrational modes of NTCDA, CuPc and PTCDA adsorbed on Ag(111) that display pronounced asymmetric or even dispersive line shapes [16, 27]; the derived values are summarized in Table 1. For NTCDA/Ag(111) and PTCDA/Ag(111) the asymmetry parameters amount to $\omega_0\tau = 0.3 - 0.4$, while notably higher values up to 1.0 are found for CuPc/Ag(111).

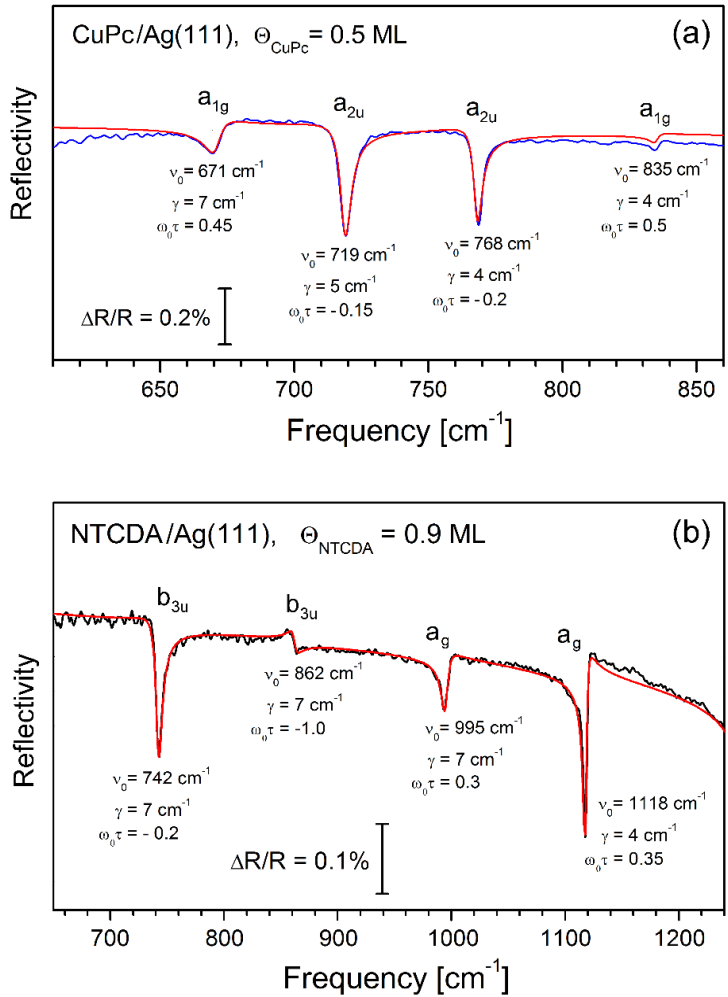

**Figure 2:**
Infrared absorption spectra of (a) CuPc/Ag(111), and (b) the relaxed NTCDA/Ag(111) monolayers. The asymmetric lines have been fitted using Fano-type line shapes according to Langreth [11]. The derived parameters are indicated in the respective panels. The indicated irreducible representations refer to the $D_{4h}$ and the $D_{2h}$ symmetry groups of CuPc and NTCDA, respectively.

We stress that $\omega_0\tau$ must not be used directly to derive the molecule - metal charge transfer time $\tau_{IDCT}$, as $\tau_{IDCT} \neq \tau$ and only for $|\mu_{mol}| \ll |\mu_{IDCT}|$ can we expect $\tau_{IDCT} \approx \tau$. This condition is actually favorably fulfilled for in-plane vibrational modes of parallel oriented planar molecules adsorbed on noble metal surfaces, used throughout in our study. We can thus conclude that $\mu_{dyn} = \mu_{mol} + \mu_{IDCT} \approx \mu_{IDCT}$, so that $\omega_0\tau_{IDCT} \approx \omega_0\tau$, i.e. the derived asymmetry parameter $\omega_0\tau$ represents a realistic measure of the phase delay $\omega_0\tau_{IDCT}$ associated with the electron transfer between molecule and Ag(111) substrate. We note that examination of out-of-plane vibrational modes is much more demanding as they come with a non-negligible $\mu_{mol}$, so that $\mu_{mol} \ll \mu_{IDCT}$ does not hold anymore; the extraction of $\tau_{IDCT}$ from observed asymmetry parameters then requires knowledge of the magnitudes of $\mu_{mol}$ and of $\mu_{IDCT}$ (actually, it is their relative ratio that is relevant).

**Table 1:**

Asymmetry parameters $\omega_0\tau$ and charge transfer times $\tau_{IDCT}$ derived from a line shape analysis of various in-plane vibrational modes of CuPc, NTCDA, and PTCDA[a] adsorbed on Ag(111).

| Molecule | Line position $[cm^{-1}]$[b] | $\omega_0\tau$ | $\tau_{IDCT}$ [fs] |
|---|---|---|---|
| CuPc | 671 | 0.45 | 3.6 |
| | 835 | 0.5 | 3.2 |
| | 1121 | 0.75 | 3.5 |
| | 1388 | 1.0 | 3.8 |
| | 1516 | 0.9 | 3.1 |
| NTCDA | 995 | 0.3 | 1.6 |
| | 1118 | 0.35 | 1.7 |
| | 1257 | 0.3 | 1.3 |
| | 1264 | 0.3 | 1.3 |
| | 1338 | 0.35 | 1.4 |
| PTCDA | 1042 | 0.3 | 1.5 |

[a] For PTCDA the number of suitable modes is limited as the majority of vibrational bands display a complex internal structure, impeding their proper evaluation

[b] $\omega_0$ is derived from vibrational mode line positions $\nu_0$ according to $\omega_0\,[1/s] = 2\pi c \nu_0\,[cm^{-1}]$

Using the above expression for $\tau_{IDCT}$, the asymmetry values in Table 1 can now be used to extract the characteristic charge transfer time $\tau_{IDCT}$, with $\omega_0$ acting as an internal clock reference. We find that the effective charge transfer times $\tau_{IDCT}$ varies notably for individual molecular species, reflecting the respective molecule-metal interaction strength. Specifically, values of about 1.5 fs are derived for NTCDA and PTCDA on Ag(111) and around 3.5 fs for CuPc/Ag(111), signaling a distinctly higher degree of non-adiabaticity in the latter case.

This marked difference in charge transfer time $\tau_{IDCT}$ is in accordance with the quite strong distortion of the initially planar structure of NTCDA and PTCDA adsorbed on Ag(111); it is primarily caused by the chemical bonding of O-atoms located at both ends of the molecules to surface Ag atoms [15, 28]. This chemical interaction is in line with the strong shift of carboxyl stretching frequencies in our IR spectra. Warping of CuPc/Ag(111), on the other hand, is rather weak [24]. The derived notably weaker molecule-substrate interaction for CuPc is in perfect agreement with the larger bonding distance of CuPc/Ag(111) (3.08 Å) [24], as compared to the values reported for the carbon backbone of NTCDA (2.997 Å) [28] and of PTCDA (2.86 Å) [15, 29].

For NTCDA on Ag(111) a correlation of the f-LUMO energy position and the respective oscillatory motion, $dE_{LUMO}/d\xi_i$, with $\xi_i$ denoting a generalized NTCDA normal mode coordinate, has been established based on DFT calculations [30]. Thereby the respective contributions to $\mu_{mol}$ (denoted as $\mu'_{nucl}$ in [30]) and $\mu_{IDCT}$ have been quantified. In fact and in accordance with our observations, out-of-plane modes ($b_{3u}$) yielded non-negligible $\mu_{mol}$, as well as opposite signs of $\mu_{mol}$ and $\mu_{IDCT}$. As a consequence, out-of-plane modes show a rather broad distribution of asymmetry parameters, consistent with the spectra in Figure 2. If we take the values for $\mu_{mol}$ and $\mu_{IDCT}$ reported in [30] for the 742 cm$^{-1}$ mode to calculate $\tau_{IDCT}$ by means of the expression above and using $\omega_0\tau = -0.2$, we obtain a value of $\tau_{IDCT} = 1.6$ fs [31]. This close match with the values derived from our analysis of in-plane modes (Table 1) is actually not surprising, as the molecular orbital associated with IDCT ought to be the same for all vibrational modes, independent of their character.

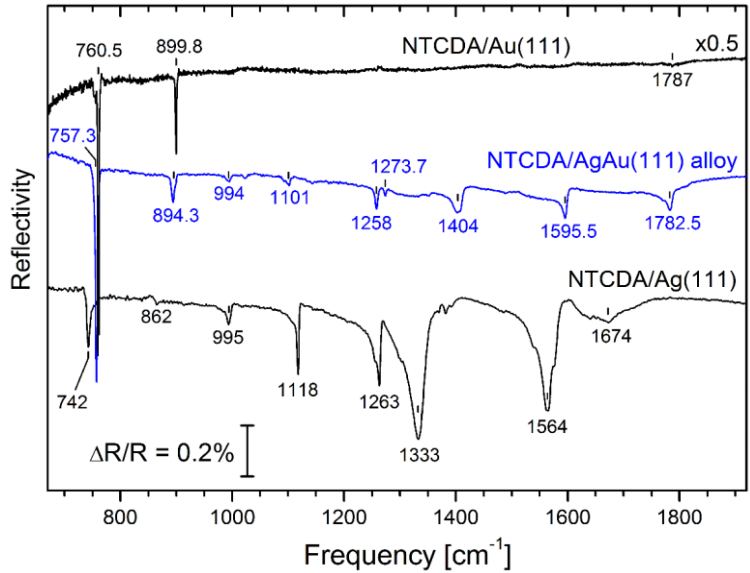

**Figure 3:**
Infrared absorption spectra of NTCDA adsorbed on Au(111) (top), Ag(111) (bottom) and an AgAu(111) surface alloy created by annealing an Ag/Au(111) monolayer to 580 K (center). Spectral resolution was 0.5 cm$^{-1}$ for NTCDA/Au(111), while 2 cm$^{-1}$ was used otherwise.

In Figure 3 we have modified the substrate electronic structure by deposition of Ag atoms on Au(111) to create an AgAu(111) surface alloy [32] after annealing to 580 K. The idea is to adjust the dynamics of vibrationally induced charge transfer processes across the molecule-metal interface. As a probe particle NTCDA has been deposited at 80 K on Au(111), the Ag/Au alloy, and Ag(111); all layers have been annealed to 400 K to induce lateral ordering and to desorb surplus second-layer species.

It is apparent that IDCT is entirely absent for NTCDA on Au(111), in contrast to NTCDA adsorbed on Ag(111), and on the AgAu(111) alloy surface. We conclude that IDCT can effectively be switched on by adding Ag atoms to the Au(111) surface. The adsorption-induced vibrational frequency shifts for NTCDA on AgAu(111) with respect to the gas phase or bulk values are relatively small and we conclude that the molecule - metal interaction, while stronger than on Au(111), is notably weaker for the alloy surface as compared to Ag(111). Most relevant in the current context, the in-plane modes of NTCDA on the Ag/Au alloy substrate display characteristic dispersive line shapes, indicating substantially larger asymmetry parameters as compared to NTCDA/Ag(111). Accordingly, the time scale $\tau_{IDCT}$ of molecule - metal charge transfer increases to about 2 fs, which is significantly slower than found for NTCDA on Ag(111). Functionalizing the metal substrate surface thus represents an effective way to tune $\tau_{IDCT}$.

**Summary**

In conclusion, we have investigated the charge transfer dynamics across the molecule-metal interface for various molecular contact layers. By means of vibrational excitations and evaluation of asymmetric line shapes induced by non-adiabatic electron - vibron coupling, the dynamic charge transport right at $\varepsilon_F$ has been probed. This is advantageous insofar as it avoids electronically excited states far above $\varepsilon_F$ that are commonly encountered in various electron spectroscopies and in this way is much more relevant to charge transport in molecular electronic devices. In our analysis the vibrational mode frequency has been used as an internal clock reference to derive the charge transfer time $\tau_{IDCT}$, representing an intrinsic property of the molecule - metal interface. Typically, $\tau_{IDCT}$ is in the few fs range, corresponding to a notable fraction of the oscillation period. Our approach to explore the dynamics of charge transfer processes across molecule-metal interfaces is both accurate and generally applicable, and represents a powerful method to characterize this technologically highly relevant material class.

**Acknowledgments**

We gratefully acknowledge discussions with U. Höfer. Funding was provided by the Deutsche Forschungsgemeinschaft (DFG, German Research Foundation), Project-ID 223848855-SFB 1083 "Structure and Dynamics of Internal Interfaces".

**References:**

[1] P. Echenique, R. Berndt, E. Chulkov, T. Fauster, A. Goldmann, and U. Höfer, Decay of electronic excitations at metal surfaces, Surface Science Reports 52, 219 (2004).

[2] F. Evers, R. Korytár, S. Tewari, and J. M. van Ruitenbeek, Advances and challenges in single-molecule electron transport, Rev. Mod. Phys. 92, 035001 (2020).

[3] Y. F. Wang, J. Kröger, R. Berndt, H. Vázquez, M. Brandbyge, and M. Paulsson, Atomic-scale control of electron transport through single molecules, Phys. Rev. Lett. 104, 176802 (2010).

[4] X.-Y. Zhu, Electron transfer at molecule-metal interfaces: A two-photon photoemission study, Ann. Rev. Phys. Chem. 53, 221 (2002).

[5] S. Hagen, Y. Luo, R. Haag, M. Wolf, and P. Tegeder, Electronic structure and electron dynamics at an organic molecule/metal interface: interface states of tetra-tertbutyl-imine/Au(111), New Journal of Physics 12, 125022 (2010).

[6] C. H. Schwalb, S. Sachs, M. Marks, A. Schöll, F. Reinert, E. Umbach, and U. Höfer, Electron lifetime in a shockley-type metal-organic interface state, Phys. Rev. Lett. 101, 146801 (2008).

[7] W. Wurth and D. Menzel, Ultrafast electron dynamics at surfaces probed by resonant Auger spectroscopy, Chem. Phys. 251, 141 (2000).

[8] A. Föhlisch, P. Feulner, F. Hennies, A. Fink, D. Menzel, D. Sanchez-Portal, P. M. Echenique, and W. Wurth, Direct observation of electron dynamics in the attosecond domain, Nature 436, 373 (2005).

[9] B. N. J. Persson and M. Persson, Vibrational lifetime for CO adsorbed on Cu(100), Solid State Commun. 36, 175 (1980).

[10] B. N. J. Persson, Vibrational Energy Relaxation at Surfaces: $O_2$ Chemisorbed on Pt(111), Chem. Phys. Lett. 139, 457 (1987).

[11] D. C. Langreth, Energy Transfer at Surfaces: Asymmetric Line Shapes and the Electron-Hole-Pair Mechanism, Phys. Rev. Lett. 54, 126 (1985).

[12] Y. J. Chabal, Electronic Damping of Hydrogen Vibration on the W(100) Surface,
Phys. Rev. Lett. 55, 845 (1985).

[13] J. E. Reutt, Y. J. Chabal, and S. B. Christman, Coupling of H vibration to substrate electronic states in Mo(100)-p(1×1)H and W(100)-p(1×1)H: Example of strong breakdown of adiabaticity, Phys. Rev. B 38, 3112 (1988).

[14] S. Soubatch, I. Kröger, C. Kumpf, and F. S. Tautz, Structure and growth of tetracene on Ag(111), Phys. Rev. B 84, 195440 (2011).

[15] A. Hauschild, K. Karki, B. C. C. Cowie, M. Rohlfing, F. S. Tautz, and M. Sokolowski, Molecular Distortions and Chemical Bonding of a Large π-Conjugated Molecule on a Metal Surface, Phys. Rev. Lett. 94, 036106 (2005).

[16] S. Thussing and P. Jakob, Structural and Vibrational Properties of CuPc/Ag(111) Ultrathin Films, J. Phys. Chem. C 120, 9904 (2016).

[17] R. Tonner, P. Rosenow, and P. Jakob, Molecular Structure and Vibrations of NTCDA Monolayers on Ag(111) from Density-Functional Theory and Infrared Absorption Spectroscopy, Phys. Chem. Chem. Phys. 18, 6316 (2016).

[18] N. L. Zaitsev, P. Jakob, and R. Tonner, Structure and vibrational properties of the PTCDA/Ag(111) interface: bilayer versus monolayer, J. Phys.: Condens. Matter 30, 354001 (2018).

[19] H. Ueba, Vibrational relaxation and pump-probe spectroscopies of adsorbates on solid surfaces, Prog. Surf. Sci. 55, 115 (1997).

[20] A. M. Wodtke, D. Matsiev, and D. J. Auerbach, Energy transfer and chemical dynamics at solid surfaces: The special role of charge transfer, Prog. Surf. Sci. 83, 167 (2008).

[21] A. Bendounan, F. Forster, A. Schöll, D. Batchelor, J. Ziroff, E. Umbach, and F. Reinert, Electronic structure of 1ML NTCDA/Ag(111) studied by photoemission spectroscopy,
Surf. Sci. 601, 4013 (2007).

[22] J. Ziroff, F. Forster, A. Schöll, P. Puschnig, and F. Reinert, Hybridization of Organic Molecular Orbitals with Substrate States at Interfaces: PTCDA on Silver, Phys. Rev. Lett. 104, 233004 (2010).

[23] Y. Zou, L. Kilian, A. Schöll, T. Schmidt, R. Fink, and E. Umbach, Chemical bonding of PTCDA on Ag surfaces and the formation of interface states, Surf. Sci. 600, 1240 (2006).

[24] I. Kröger, B. Stadtmüller, C. Stadler, J. Ziroff, M. Kochler, A. Stahl, F. Pollinger, T.-L. Lee,
J. Zegenhagen, F. Reinert, and et al., Submonolayer Growth of Copper-Phthalocyanine on Ag(111), New J. Phys. 12, 083038 (2010).

[25] J. Ziroff, P. Gold, A. Bendounan, F. Forster, and F. Reinert, Adsorption energy and geometry of physisorbed organic molecules on Au(111) probed by surface-state photoemission,
Surf. Sci. 603, 354 (2009).

[26] S. Soubatch, C. Weiss, R. Temirov, and F. S. Tautz, Site-Specific Polarization Screening in Organic Thin Films, Phys. Rev. Lett. 102, 177405 (2009).

[27] C. R. Braatz, G. Öhl, and P. Jakob, Vibrational Properties of the Compressed and the Relaxed 1,4,5,8- Naphthalene-Tetracarboxylic Dianhydride Monolayer on Ag(111),
J. Chem. Phys. 136, 134706 (2012).

[28] C. Stadler, S. Hansen, A. Schöll, T.-L. Lee, J. Zegenhagen, C. Kumpf, and E. Umbach, Molecular Distortion of NTCDA Upon Adsorption on Ag(111): A Normal Incidence X-Ray Standing Wave Study, New J. Phys. 9, 50 (2007).

[29] A. Hauschild, R. Temirov, S. Soubatch, O. Bauer, A. Schöll, B. C. C. Cowie, T.-L. Lee, F. S. Tautz, and M. Sokolowski, Normal-incidence x-ray standing-wave determination of the adsorption geometry of PTCDA on Ag(111): Comparison of the ordered room-temperature and disordered low-temperature phases, Phys. Rev. B 81, 125432 (2010).

[30] P. Rosenow, P. Jakob, and R. Tonner, Electron-Vibron Coupling at Metal-Organic Interfaces from Theory and Experiment, J. Phys. Chem. Lett. 7, 1422 (2016).

[31] For the 862 cm$^{-1}$ mode in Figure 2b, a similar analysis lead to unreasonable $\tau_{IDCT}$ values, which is ascribed to a nearly complete compensation of $\mu_{mol}$ and $\mu_{1,\,IDCT}$.

[32] H. Cercellier, C. Didiot, Y. Fagot-Revurat, B. Kierren, L. Moreau, D. Malterre, and F. Reinert, Interplay between structural, chemical, and spectroscopic properties of Ag/Au(111) epitaxial ultrathin films: A way to tune the Rashba coupling, Phys. Rev. B 73, 195413 (2006).